\documentclass[10pt, conference]{IEEEtran}
\usepackage{color}
\usepackage{cite}
\ifCLASSINFOpdf
  \usepackage[pdftex]{graphicx}
\else
\fi
\usepackage{amsmath}
\usepackage{amssymb}
\usepackage{algorithm}
\usepackage{algpseudocode}
\usepackage{subcaption}
\usepackage[utf8]{inputenc}
\usepackage[english]{babel}
\usepackage[section]{placeins}
\usepackage{subcaption}
\usepackage{authblk}
\newtheorem{theorem}{Theorem}[section]

\newtheorem{lemma}[theorem]{Lemma}

\usepackage{cuted}
\usepackage{flushend}

\begin{document}

	\title{Age of Information-Aware Cognitive Shared Access Networks with Energy Harvesting
    }
\author[1]{Georgios Smpokos}
\author[2]{Dionysis Xenakis}
\author[3]{Marios Kountouris}
\author[1]{Nikolaos Pappas}
\affil[1]{Link\"oping University, Sweden}
\affil[2]{National and Kapodistrian University of Athens, Greece}
\affil[3]{University of Granada, Spain}

\maketitle

\begin{abstract}
This study investigates a cognitive shared access network with energy harvesting capabilities operating under Age of Information (AoI) constraints for the primary user. Secondary transmitters are spatially distributed according to a homogeneous Poisson Point Process (PPP), while the primary user is located at a fixed position. The primary transmitter handles bursty packet arrivals, whereas secondary users operate under saturated traffic conditions. To manage interference and energy, two distinct zones are introduced: an energy harvesting zone around the primary transmitter and a guard zone around the primary receiver, within which secondary transmissions are prohibited. Secondary users access the channel probabilistically, with access decisions depending on their current battery state (charged or empty) and their location relative to the guard zone. Our objective is to analyze the primary user’s AoI performance under three distinct packet management policies.


\end{abstract}

\IEEEpeerreviewmaketitle

\section{Introduction}
Age of Information (AoI) has emerged as a critical performance metric in shared access network environments, quantifying the timeliness (freshness) of the received data \cite{Kosta1}. Introduced in foundational studies \cite{Altman1, Kaul1, Kaul2}, AoI measures the time elapsed from the time a timestamped status update is generated at its source until it is successfully received and decoded at its destination. Maintaining a low average AoI ensures that the delivered information remains up-to-date and relevant. The concept has since broadened to incorporate update costs, information-value metrics, and non-linear AoI evaluations~\cite{Kosta2, Sun1, Sun2, Yates2}. The importance of preserving information freshness, as captured by AoI, spans a wide range of application domains, including unmanned aerial vehicle (UAV)-based Internet of Things (IoT) networks~\cite{Elmagid1,Elmagid2,Mankar1}, multicast and ad hoc wireless communications~ \cite{Kadota1,Hsu1,Kadota2,Buyukates1,He1,Salimnejad1}, wireless sensor networks for environmental and healthcare monitoring, data warehousing systems \cite{Wu1,Arafa1,Arafa2,Nath1, Chen2,Gu1,Fountoulakis1}, and web caching technologies \cite{Yu1,Kam1,Yates1,Zhong1}.
Energy Harvesting (EH) enables devices, such as sensors, to recharge their batteries by capturing electromagnetic energy, making EH particularly suitable for deployment in environments where battery replacement or a wired power supply is impractical. Another relevant use case for EH arises in device-to-device (D2D) communication systems within cellular networks, where D2D users harvest energy from cellular transmissions. Performance evaluation and optimization of such systems have been studied in \cite{Yang1,Sakr1}, where secondary nodes access the medium under specific constraints and primary nodes are typically assumed to operate under saturated traffic. These works analyze the joint impact of random packet arrivals at primary nodes and opportunistic EH-enabled channel access by secondary users, with average delay as a key performance metric \cite{Chen1}. Additional studies, such as \cite{Pappas2, Jeon1, Pappas3, Xu1}, apply queueing theory to investigate EH-enabled networks and quantify how energy harvesting influences the stability region in small-scale systems.

\section{System Model}
A primary transmitter (PT) is located at the center of a circular, omnidirectional coverage area modeled as a disk $\mathcal{C}$ of radius $R$. The associated primary receiver (PR) is fixed at $(0,y_0)$, at distance $d_p$ from the PT. Secondary transmitters (STs) are spatially distributed over $\mathcal{C}$ according to a homogeneous Poisson point process (PPP) $\Phi_s$ with intensity $\lambda_s$. Each ST is paired with a designated secondary receiver (SR) placed at a fixed distance $d_s$ in a uniformly random direction from its corresponding ST.
Each ST is equipped with a radio frequency (RF) energy harvesting module that converts ambient RF energy into direct current (DC) power for battery storage. A key notion in EH systems is the \emph{energy harvesting zone}: a disk of radius $r_{\mathrm{eh}}$ centered at an RF source within which the received power exceeds the minimum conversion threshold. Any ST located inside this zone can accumulate sufficient energy within a single time slot \cite{Huang1,Lee2,Yang2}.

In each time slot, the subset of STs that can be fully charged by the primary transmitter's (PT's) RF transmission is given by:
\begin{equation}
\mathcal{S} = \Phi_s \cap \mathcal{C}_{eh}(0, r_{eh}),
\end{equation}
where $\mathcal{C}(a, b)$ denotes a disk centered at point $a$ with radius $b$. In this context, $\mathcal{C}_{eh}(0, r_{eh})$ represents the energy harvesting zone centered at the PT, while $\Phi_s$ denotes the spatial distribution of STs, modeled as a homogeneous PPP with intensity $\lambda_s$.

We assume that the primary transmitter operates at power level $P_p$, while each secondary transmitter transmits at a lower power level $P_s$, where $P_s \ll P_p$. This power disparity reflects the extended transmission range of the PT and the limited communication distances of the STs. Although the STs operate at reduced individual power levels, their simultaneous transmissions can produce substantial cumulative interference at the PR, thereby degrading the performance of the PT–PR link. To mitigate this interference, a protection zone is defined around the PR, prohibiting ST transmissions within a disk of radius $r_{gz}$ centered at the PR’s location $(0, y_0)$~\cite{Hasan1}.

Incorporating this interference-aware constraint, the set of STs eligible to transmit in a given time slot must satisfy two conditions:
\begin{enumerate}
    \item The ST must reside within the EH zone and have successfully harvested sufficient energy, i.e., it must belong to the set $\mathcal{S}$;
    \item The ST must lie outside the guard zone surrounding the PR.
\end{enumerate}

Accordingly, the set of active STs in a given time slot is defined as:
\begin{equation}
\mathcal{S}_A = \mathcal{S} \setminus \left( \Phi_s \cap \mathcal{C}_{gz}(y_0, r_{gz}) \right),
\end{equation}
where $\mathcal{C}_{gz}(y_0, r_{gz})$ denotes the guard zone centered at the PR’s location $(0, y_0)$ with radius $r_{gz}$.

For the network configuration under consideration, we define the following probabilities related to the behavior and status of STs:
\begin{itemize}
    \item $p_{tr}$: the probability that an ST transmits in a given time slot,
    \item $p_{ch}$: the probability that an ST's battery is fully charged at the beginning of a time slot,
    \item $p_{eh}$: the probability that an ST is located within the EH zone of the PT,
    \item $p_{gz}$: the probability that an ST is located within the guard zone of the PR and is therefore prohibited from transmitting.
\end{itemize}

We consider a dynamic environment in which STs are mobile across time slots. Consequently, the probability that an ST lies outside the guard zone and the probability that it was fully charged in the previous time slot are treated as independent events. According to the adopted random access protocol, an eligible ST attempts transmission with probability $p_s$ in any given time slot.

Thus, the overall probability that an ST becomes active, i.e., transmits in a given time slot, is given by:
\begin{equation}
    p_{tr} = p_{ch} \cdot (1 - p_{gz}) \cdot p_s.
\end{equation}

This expression represents the proportion of active ST transmissions in each network configuration, capturing the combined effects of energy availability, the spatial restrictions imposed by the guard zone, and the probabilistic nature of access control.

The ability of the PR to successfully decode signals from the PT The ability of the PR to successfully decode signals from the PT depends on the signal-to-interference-plus-noise ratio (SINR) measured at the PR. This SINR accounts for the cumulative interference generated by the STs that are simultaneously active during the same time slot. It is expressed as:
\begin{equation}
\label{eq_SINR}
\text{SINR}p = \frac{P_p |h_P|^2 d_p^{-\alpha}}{\sigma^2 + \sum{i \in \mathcal{S}{\text{tr}}} P_s |h_i|^2 d_i^{-\alpha}},
\end{equation}
where $\mathcal{S}{\text{tr}} \subset \mathcal{S}_A$ denotes the set of STs transmitting in the current time slot. The term $|h_P|^2$ represents the small-scale Rayleigh fading power gain on the PT–PR link, and $d_p$ is the distance between the PT and PR. Similarly, $|h_i|^2$ denotes the fading power gain between the $i$-th ST and the PR, with $d_i$ as the corresponding distance. The parameter $\alpha > 2$ is the path loss exponent, and $\sigma^2$ denotes the background thermal noise power, modeled as additive white Gaussian noise (AWGN).

This expression captures the impact of concurrent ST transmissions on the SINR at the PR and, consequently, on the reliability of the primary communication link.

\section{Performance Analysis}
For secondary users, a key metric reflecting overall network performance is the \emph{average secondary throughput}. To evaluate this metric, two fundamental probabilities must first be defined: the \emph{active probability}, which denotes the likelihood that a secondary user is ready to transmit in a given time slot, and the \emph{successful transmission probability}, which represents the likelihood that a secondary user's signal is successfully decoded at its intended receiver.

\begin{figure}[!htb]
\centering
\includegraphics[width=2.6in]{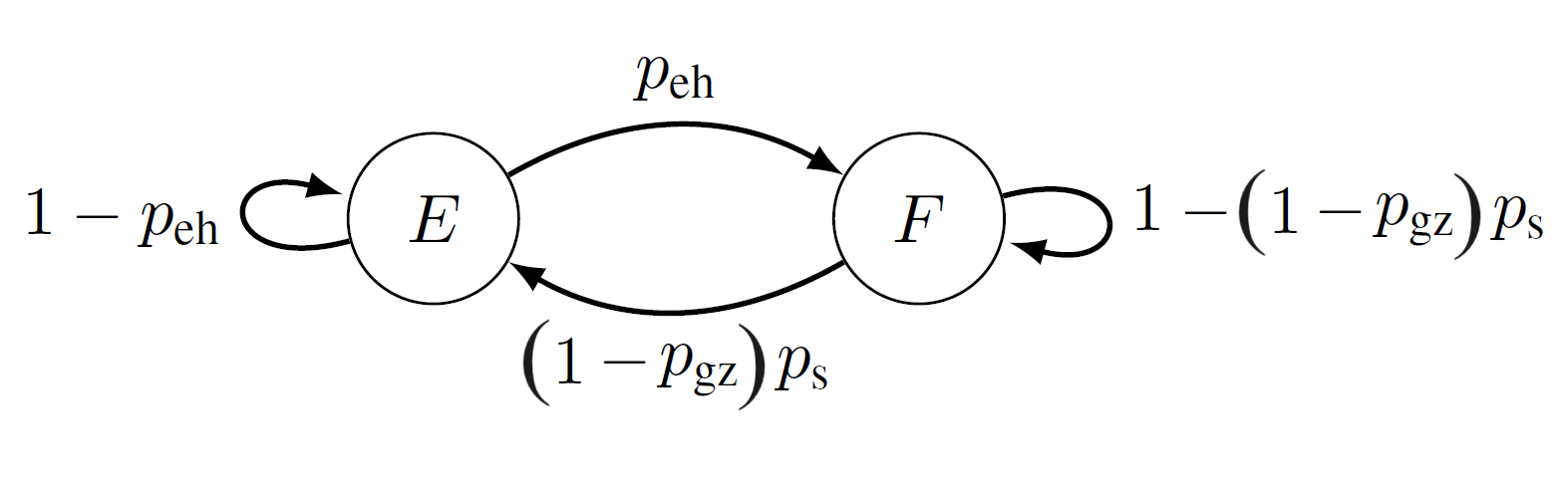}
\captionsetup{font = small}
\caption{Two-state Discrete-Time Markov Chain (DTMC) modeling the ST battery state evolution.}
\label{fig2}
\end{figure} 

In our analysis, we employ a single-slot charging and discharging model, where each ST's battery is represented as a two-state discrete-time Markov chain (DTMC), as depicted in Figure~\ref{fig2}. In this model, state \textit{F} corresponds to a fully charged battery, allowing the ST to transmit, whereas state \textit{E} denotes an empty battery, preventing transmission.

Using this DTMC approach, we compute the stationary probability that a secondary user has sufficient energy to transmit, which constitutes a key component in evaluating the overall average secondary throughput.

The probability that an ST is fully charged and thus able to transmit in the next time slot is given by:
\begin{equation}
    p_{ch} = \frac{p_{eh}}{p_{eh} + p_s - p_{gz} p_s}.
\end{equation}
The probabilities that an ST is located within the EH zone and within the PR guard zone are derived based on geometric considerations and the law of cosines, respectively:
\begin{equation}
    p_{eh} = \frac{r_{eh}^2}{R^2},
\end{equation}
\begin{equation}
    p_{gz} =
    \begin{cases}
        \dfrac{r_{gz}^2}{R^2}, & r_{gz} \leq R - d_p, \\[6pt]
        \dfrac{\phi r_{gz}^2}{\pi R^2} + \dfrac{\varphi}{\pi} - \dfrac{d_p}{\pi R} \sin{\varphi}, & r_{gz} > R - d_p,
    \end{cases}
\end{equation}
where $\phi = \arccos\!\left(\frac{d_p^2 + r_{gz}^2 - R^2}{2 d_p r_{gz}}\right)$ and $\varphi = \arccos\!\left(\frac{R^2 + d_p^2 - r_{gz}^2}{2 d_p R}\right)$.

In our analysis, we define the secondary user throughput as the average number of successful packet transmissions per time slot per unit area, measured in packets/slot/$\text{m}^2$. Let $\overline{T}_s$ denote the average secondary throughput, expressed as  
\begin{equation}
    \overline{T}_s = \lambda_{s} p_{tr} p_{sx} = \lambda_s \frac{p_{eh}}{p_{eh} + p_s - p_{gz} p_s} (1 - p_{gz}) p_s p_{sx},
\end{equation}
where $p_{sx}$ denotes the probability of a successful transmission from an ST to its designated SR. To evaluate $p_{sx}$, we first analyze the SINR, assuming that a transmission is deemed successful if the received SINR exceeds a predefined threshold $\theta$. Considering that the SR of the $i$-th secondary transmitter–receiver pair is located at the origin, we can derive the following expression:

\begin{equation}
    \text{SINR}_i=\frac{P_s|h_{i,i}|^2{d_s}^{-\alpha}}{\sigma^2+\sum_{j\in \mathcal{S_{\text{tr}}}\setminus \{ i \} } P_s |h_{j,i}|^2 d_{j,i}^{-\alpha}+P_p|h_{p,i}|^2 d_{p,i}^{-\alpha}},
\end{equation}
where $\mathcal{S}_{\text{tr}}$ denotes the set of active STs, $|h_{j,i}|^2$ represents the small-scale Rayleigh fading gain from transmitter $j$ to receiver $i$, assumed to have unit mean, and $d_{j,i}$ is the distance between transmitter $j$ and the SR $i$. As in equation~(\ref{eq_SINR}), the term $d^{-\alpha}$ models the distance-dependent path-loss attenuation, where $\alpha > 2$ is the path-loss exponent, and $\sigma^2$ denotes the thermal noise power.

Due to the presence of a guard zone around the PR, the interference field experienced by SRs becomes non-isotropic. Consequently, STs located near the boundary of the guard zone tend to experience higher transmission success probabilities because of the reduced interference in that region.

\begin{equation}
\label{eq_psx}
\begin{split}
    p_{sx} & =\mathbb{P}[\text{SINR}_i > \theta] \\ 
    & \overset{(a)}= \text{exp} (-\theta d_s^\alpha I_n ) \mathop{\mathbb{E}_{d_{p,i}}}  \left[ \frac{1}{1+ \frac{\theta P_p d_s^a}{P_s d_{p,i}^a}} \right] \text{exp} \left( - \frac{\theta \sigma^2 d_s^\alpha}{P_s} \right) \\ 
    & \overset{(b)}\simeq \mathcal{L}_{I_n} (d_s^a I_n) \frac{\text{exp} \left( - \frac{\theta \sigma^2 d_s^\alpha}{P_s} \right)}{1 + \frac{d_s^2} {\mathbb{E}[d_{p,i}]^2} \left( \theta \frac{P_p}{P_s} \right)^{\frac{2}{a}}} \\
    & \overset{(c)}\approx \text{exp} \left( - \pi \lambda_s p_{ch} p_s d_s^2 \theta^{\frac{2}{a}} \right) \frac{\text{exp} \left( - \frac{\theta \sigma^2 d_s^a}{P_s} \right)}{1 + \frac{d_s^2} {\mathbb{E}[d_{p,i}]^2} \left( \theta \frac{P_p}{P_s} \right)^{\frac{2}{a}}},
\end{split}
\end{equation}
where $I_n = \sum\limits_{j \in \mathcal{S}_{\text{tr}}} |h_{j,i}|^2 d_{j,i}^{-\alpha}$ denotes the normalized aggregate interference at the SR, and $\mathcal{L}_{I_n}(s) = \mathbb{E} \left[ e^{-s I_n} \right]$ represents the Laplace transform of the interference. In equation~(\ref{eq_psx}), step~(a) follows from the exponential probability density function of $|h_{i,i}|^2$; step~(b) results from applying the Laplace transform of the interference and the approximation introduced in~\cite{Lee3}; and step~(c) approximates $p_{sx}$ using the probability generating functional (p.g.fl) of a PPP. Given the random placement of STs across the network coverage area, each SR is uniformly distributed within the disk $\mathcal{C}(0,R)$. The expected distance between the PT and an SR is thus given by $\mathbb{E}[d_{p,i}] = \frac{2R}{3}$.

In this part of the performance evaluation, we analyze the primary user's performance in terms of its \emph{average AoI} under three distinct packet management strategies. These strategies define how status update packets are handled at the PT, particularly under varying packet arrival and service probabilities. The considered scenarios are as follows:

\begin{itemize}
    \item \textit{First-Come First-Served (FCFS)}: A standard queueing model without packet management, where packets are served strictly in the order of their arrival.
    \item \textit{Queue with Replacement (QR)}: A single-packet buffer in which any newly arrived packet replaces the one currently stored (if any), ensuring that only the most recent status update is retained for transmission.
    \item \textit{Generate-at-Will (GW)}: A model in which status update packets are generated only when the channel is immediately available for transmission; if the transmission attempt fails, the packet is discarded.
\end{itemize}

In the considered setup, $\mu_p$ denotes the success probability of a primary packet transmission within a given time slot. Using the SINR definition in equation~(\ref{eq_SINR}) in Section~II, and for a given threshold $\theta$, we proceed analogously to the analysis in Section~III-A for the secondary user success probability. This yields the PT’s service rate as follows:

\begin{equation}
\label{eq_mu_p}
\begin{split}
    \mu_p & =\mathbb{P}[\text{SINR}_p > \theta] \\ 
    & = \text{exp} \left( -\frac{\theta d_p^\alpha P_s}{P_p} I_s \right) \text{exp}  \left( -\frac{\theta \sigma^2 d_p^\alpha}{P_p} \right) \\ 
    & = \mathcal{L}_{I_s} \left(\frac{\theta d_p^\alpha P_s}{P_p} \right) \text{exp} \left( -\frac{\theta \sigma^2 d_p^\alpha}{P_p} \right) \\
    & = \mathcal{L}_{I}^* \left(\frac{\theta d_p^\alpha P_s}{P_p},\lambda_s p_{tr}, r_{gz}  \right) \text{exp} \left( -\frac{\theta \sigma^2 d_p^\alpha}{P_p} \right),
\end{split}
\end{equation}
where $I_s = \sum\limits_{i \in \mathcal{S}_{\text{tr}}} P_s |h_i|^2 d_i^{-\alpha}$ denotes the aggregate interference from secondary users at the PR, and $\mathcal{L}_{I}^*(t, \lambda_a, r)$ is defined as in~\cite{Chen3} and represents the Laplace transform of the interference, given by
\[
    \mathcal{L}_{I}^*(t, \lambda_a, r) = \exp\!\left(-2\pi \lambda_a \int_{r}^{\infty} \frac{t \upsilon^{-\alpha}}{1 + t \upsilon^{-\alpha}} \upsilon \, d\upsilon \right).
\]

\begin{figure}[!htb]
\centering
\includegraphics[width=2.8in]{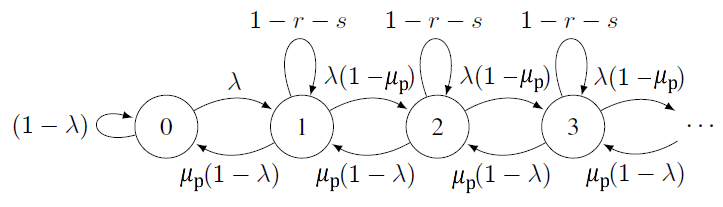}
\captionsetup{font = small}
\caption{DTMC representing the evolution of the \textit{Geo/Geo/1} queue at the PT.}
\label{fig3}
\end{figure}

\subsubsection{\textbf{FCFS - Geo/Geo/1 Queue}}
The queue at the PT is modeled as a DTMC, illustrated in Figure~\ref{fig3}, where $\lambda$ denotes the packet arrival rate and $\mu_p$ represents the service rate. Each state in the DTMC corresponds to the number of packets in the PT’s queue.

\begin{lemma}
From the DTMC presented in Figure~\ref{fig3}, the steady-state probabilities are given by
\begin{equation}
    \pi_n = \rho^{n-1} \pi_1, \quad n \geq 1,
\end{equation}
\begin{equation}
    \pi_0 = \frac{\mu_p (1 - \lambda)}{\lambda} \pi_1,
\end{equation}
where $\rho = \frac{\lambda (1 - \mu_p)}{\mu_p (1 - \lambda)}$, $\pi_1 = \frac{\lambda (1 - \rho)}{\mu_p}$, $r = \lambda (1 - \mu_p)$, $s = \mu_p (1 - \lambda)$, and $\mu_p$ is given in equation~(\ref{eq_mu_p}).
\end{lemma}

The queue at the PT is stable as long as the arrival rate satisfies $\lambda < \mu_p$~\cite{Loynes1}. Following the analysis presented in~\cite{Kosta3}, the average AoI $\overline{\Delta}_p$ is expressed as
\begin{equation}
    \overline{\Delta}_p = \frac{1}{\lambda} + \frac{1 - \lambda}{\mu_p - \lambda} - \frac{\lambda}{\mu_p^2} + \frac{\lambda}{\mu_p}.
\end{equation}

\subsubsection{\textbf{Queue with Replacement (QR)}}
The second packet management strategy considered is the \emph{queue with replacement}, in which newly arriving packets replace any existing packet waiting in the queue, provided it has not yet been transmitted. This model assumes a single-packet buffer at the PT, allowing at most one packet to be held in addition to the one currently being served. If a new status update arrives while the buffer is occupied and the queued packet has not yet been transmitted, the existing packet is discarded and replaced by the new arrival. The queue with a replacement policy at the PT can be represented using a three-state DTMC~\cite{Costa3}. The states are defined as follows:

\begin{enumerate}
    \item The system is empty (no packets are present at the PT).
    \item One packet is currently in service.
    \item One packet is in service and another is waiting in the queue (the buffer is full).
\end{enumerate}

Although this policy does not alter the maximum number of packets that can reside in the system at any given time, it allows newly arriving updates to overwrite older ones still waiting in the queue. Consequently, it prioritizes information freshness over delivery reliability.

\begin{figure}[!htb]
\centering
\includegraphics[width=2.1in]{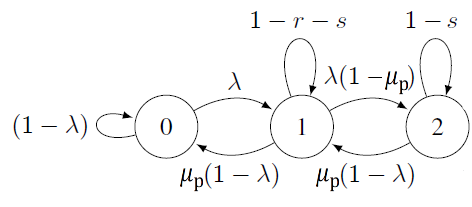}
\captionsetup{font = small}
\caption{DTMC modeling the queue evolution at the PT under the replacement policy. Here, $r = \lambda (1 - \mu_p)$ and $s = \mu_p (1 - \lambda)$.}
\label{fig4}
\end{figure}

\begin{lemma}
From the DTMC presented in Figure~\ref{fig4}, the steady-state probabilities are given by
\begin{equation}
    \pi_n = \frac{\lambda^n (1 - \mu_p)^{n-1}}{\mu_p^n (1 - \lambda)^n} \pi_0, \quad n \in \{1, 2\},
\end{equation}
\begin{equation}
    \pi_0 = \frac{\lambda - \mu_p}{\lambda \rho^2 - \mu_p},
\end{equation}
where $\rho = \frac{\lambda (1 - \mu_p)}{\mu_p (1 - \lambda)}$, $r = \lambda (1 - \mu_p)$, $s = \mu_p (1 - \lambda)$, and $\mu_p$ is given in equation~(\ref{eq_mu_p}).
\end{lemma}

We evaluate the probability that a packet in the PT’s queue is dropped based on the steady-state probabilities derived from the DTMC in Figure~\ref{fig4}, using Lemma~2, as follows:
\begin{equation}
\label{eq_pd}
\begin{split}
    p_d  & =  \pi_1 \lambda(1-\mu_p) + \pi_2 (1-s)\\ 
    & = \frac{\lambda^2(1-\mu_p)}{\mu_p^2(1-\lambda)}\biggl(1+\frac{\lambda}{\mu_p(1-\lambda)}+\frac{\lambda^2(1-\mu_p)}{\mu_p^2(1-\lambda)}\biggl)^{-1}.  
\end{split}
\end{equation}
The effective arrival rate, accounting for the packet drop probability, is expressed as follows:
\begin{equation}
\label{eq_leffective}
\begin{split}
    \lambda_e & = \lambda (1- p_d)\\ 
    & = \lambda - \frac{\lambda^3(1-\mu_p)}{\lambda^2(1-\mu_p) + \lambda(1-\mu_p)\mu_p+\mu_p^2}. 
\end{split}
\end{equation}
To evaluate the average AoI of the PT under the queue-with-replacement policy, we follow the analysis in~\cite{Kosta3}:
\begin{equation}
\label{eq_aoi}
\begin{split}
    \overline{\Delta}_p & =  \frac{1}{\delta} \Biggl( \lambda\mu_p\epsilon \biggl( \frac{\delta}{2\lambda \mu_p \epsilon} + \frac{\lambda\Bigl(\lambda(3\mu_p-2)-2\mu_p+1 \Bigl)}{\lambda^2(\mu_p-1)^2+\lambda\mu_p(1-2\mu_p)+\mu_p^2} \\
     & + \frac{\lambda^3(\mu_p-2)(\mu_p-1)+\lambda^2(\mu_p-2)(\mu_p-1)\mu_p}{2\lambda^2 \mu_p^2 \epsilon} \\
     & +\frac{\lambda \mu_p^2(2-3\mu_p)+2\mu_p^3}{2\lambda^2 \mu_p^2 \epsilon}
      + \frac{1-\lambda}{\lambda \mu_p} + \frac{2\lambda+1}{\epsilon}\\
     & -\frac{\lambda+1}{\epsilon^2}+\frac{1}{\mu_p^2} \biggl) \Biggl) 
\end{split}
\end{equation}
where $\delta = \lambda^2(1-\mu_p)+\lambda(1-\mu_p)\mu_p+\mu_p^2$ and $\epsilon = \lambda+\mu_p - \lambda \mu_p$.
\\
\subsubsection{\textbf{Generate-at-Will policy (GW)}}
In this subsection, we derive the expression for the average AoI under the \textit{generate-at-will} packet management strategy. In this scheme, the PT generates and transmits a status update only when the channel is available. If the channel is busy, the update is discarded, and a new one is generated in the next time slot according to the sampling process.

The status update generation follows a Bernoulli process with success probability $\lambda$, which also represents the sampling rate of the PT. This rate corresponds to the frequency at which updates are generated under this policy.

The average AoI at the primary receiver under the \textit{generate-at-will} policy is expressed as follows:

\begin{equation}
\label{eq_gw}
    \overline{\Delta}_p = \frac{1}{\mu_p q},
\end{equation}
where $\mu_p$ denotes the service rate (i.e., the probability of a successful transmission in a time slot), and $q$ represents the effective packet generation (or sampling) rate at the PT.

This model captures the trade-off between aggressive sampling, which aims to maintain a low AoI, and the increased risk of packet drops due to channel unavailability. 

\section{Numerical Analysis}

\begin{figure}[!htb]
\centering
\includegraphics[width=2.8in]{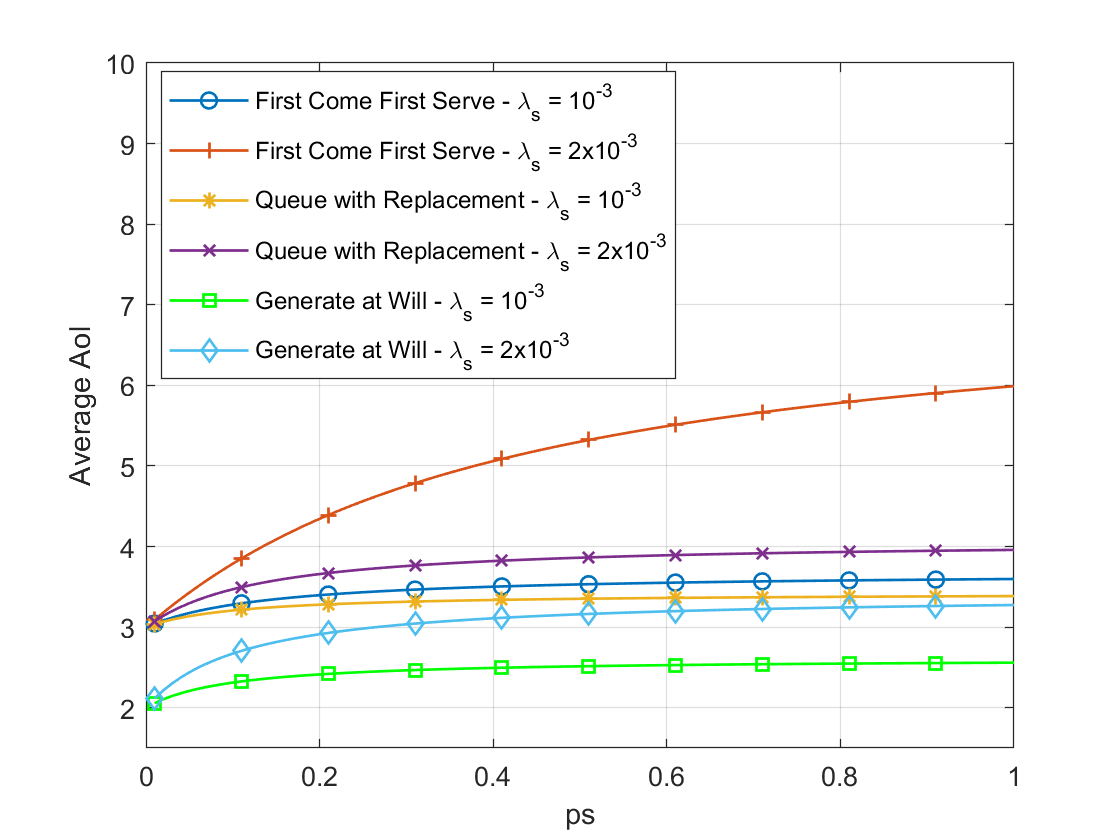}
\captionsetup{font = small}
\caption{Average AoI comparison among FCFS, QR, and GW as a function of the access probability $p_s$.}
\label{fig10}
\end{figure}

In this section, we evaluate the AoI performance for all three packet management strategies by varying multiple system parameters. Figure~\ref{fig10} compares the AoI under two different secondary user density scenarios ($\lambda_s = 10^{-3}$ and $2 \times 10^{-3}$ points/m\textsuperscript{2}) as a function of the access probability, with fixed values of $r_{eh} = 80$~m and $r_{gz} = 120$~m. Among the three strategies, the GW policy consistently achieves the lowest AoI.

When comparing FCFS and QR, the latter achieves better AoI performance as the access probability $p_s$ of the secondary users increases.

\begin{figure}[!htb]
\centering
\begin{subfigure}[b]{0.48\linewidth}
\centering
\includegraphics[width=\linewidth]{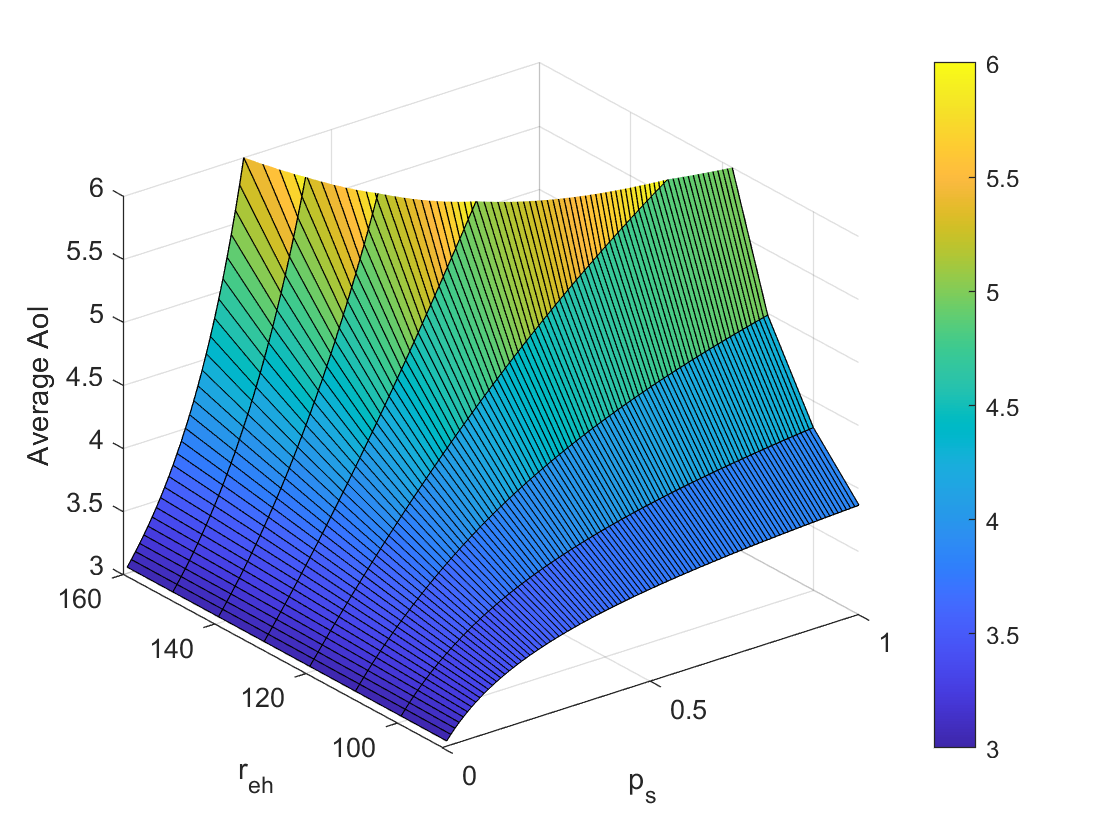}
\caption{First Come First Serve queuing policy.}
\includegraphics[width=\linewidth]{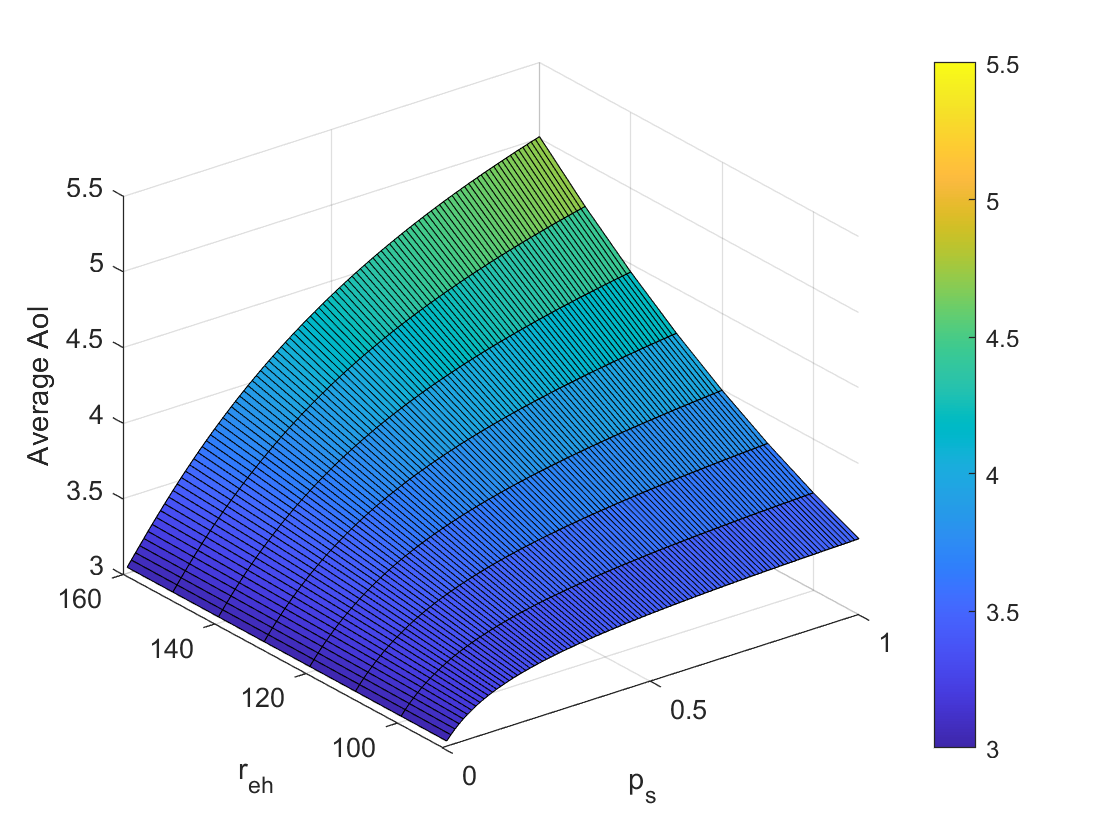}
\caption{Queue with Replacement queuing policy.}
\includegraphics[width=\linewidth]{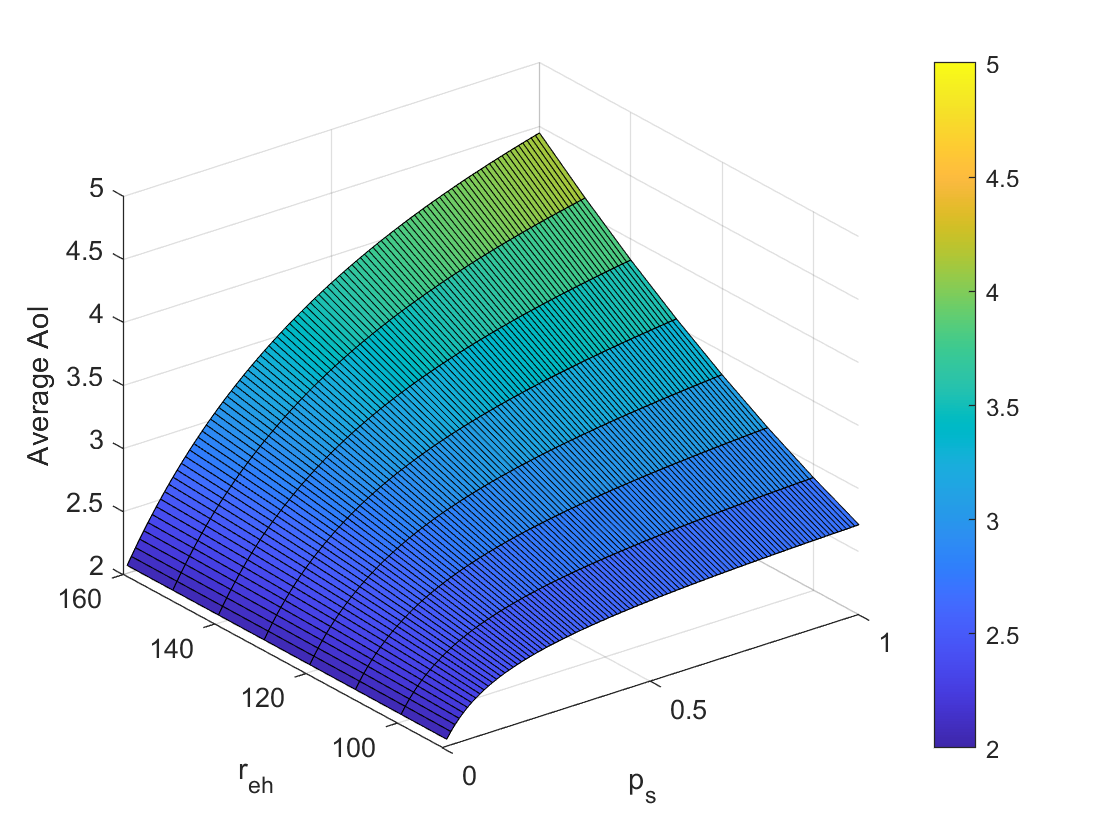}
\caption{Generate-at-Will policy.}
\end{subfigure}
\hfill
\begin{subfigure}[b]{0.48\linewidth}
\centering
\includegraphics[width=\linewidth]{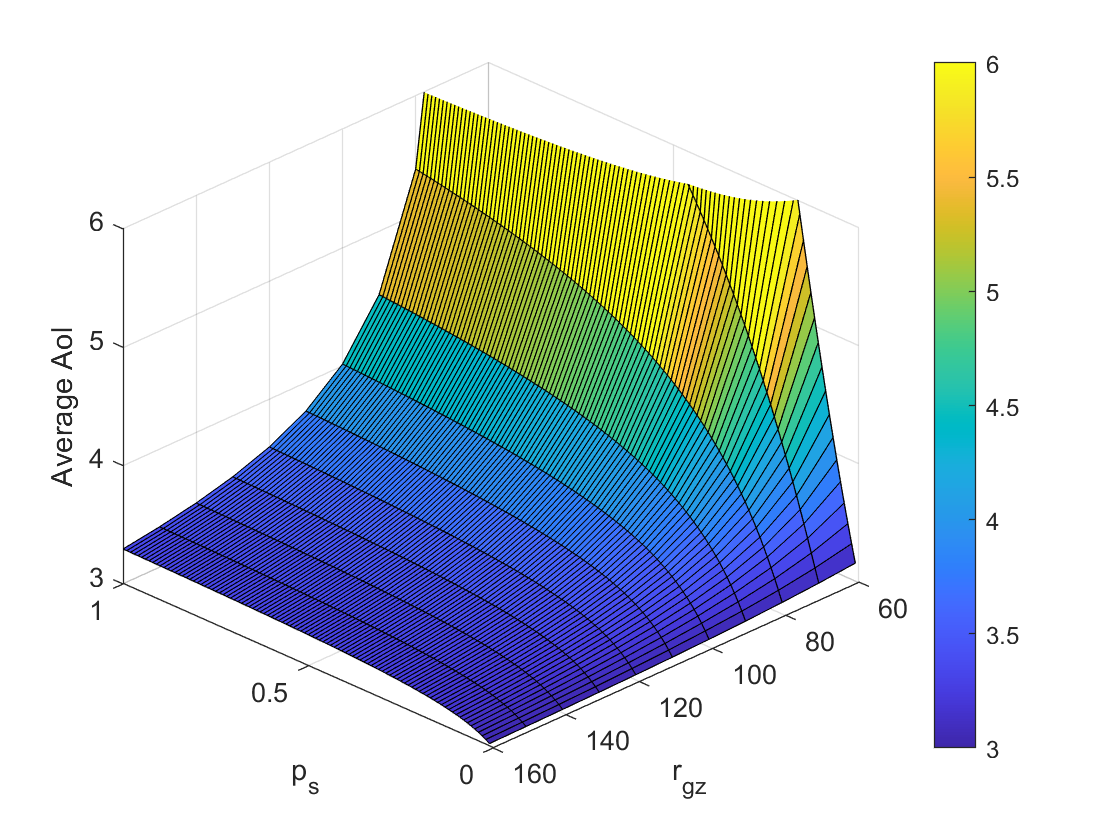}
\caption{First Come First Serve queuing policy.}
\includegraphics[width=\linewidth]{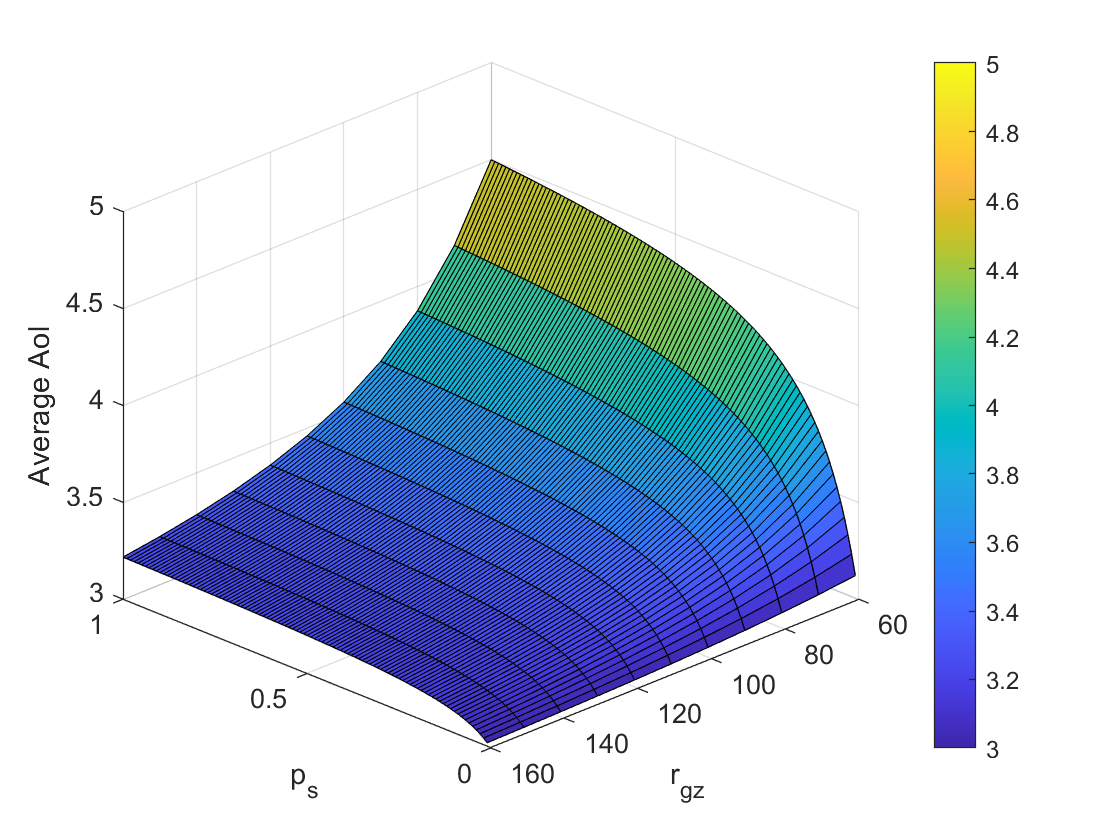}
\caption{Queue with Replacement queuing policy.}
\includegraphics[width=\linewidth]{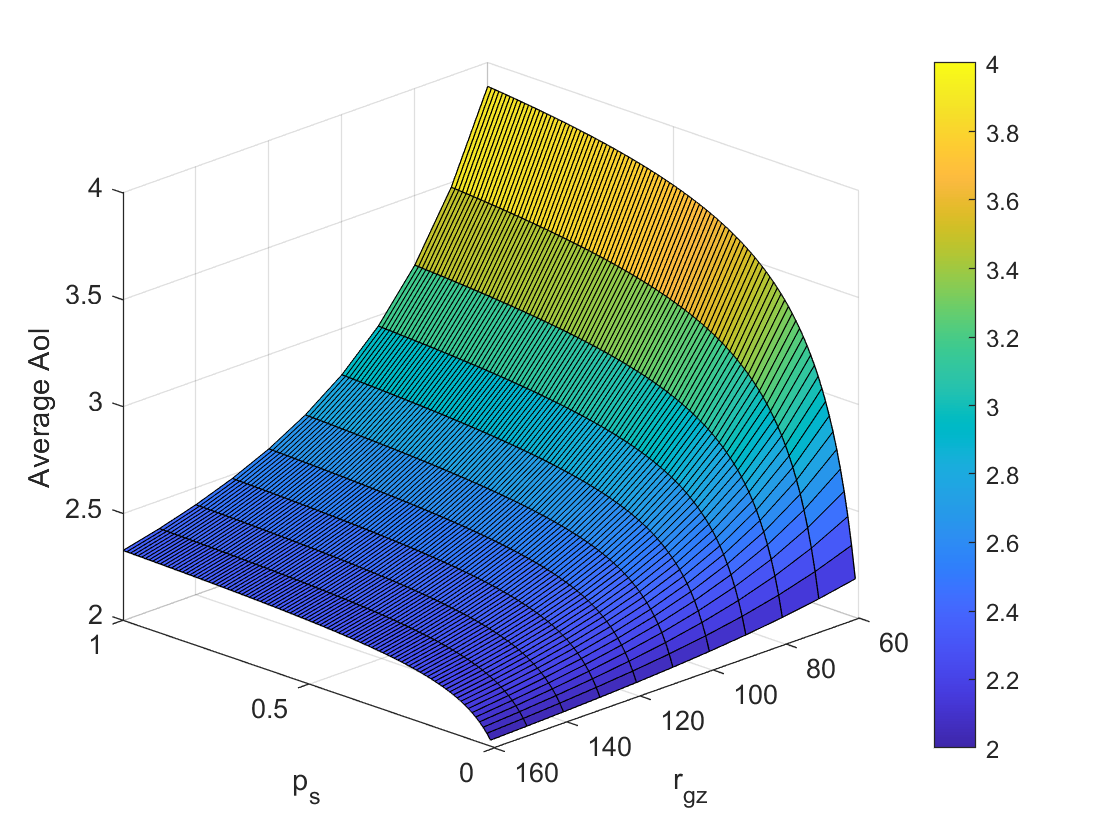}
\caption{Generate-at-Will policy.}
\end{subfigure}

\caption{Average AoI behavior of the considered policies. Left: as a function of the access probability $p_s$ and the energy harvesting zone radius $r_{eh}$. Right: as a function of the access probability $p_s$ and the guard zone radius $r_{gz}$.}
\label{fig:combined_aoi}
\captionsetup{font = small}
\end{figure}

Figure~\ref{fig:combined_aoi} presents 3D surface plots of the average AoI performance for the three queueing strategies, evaluated as functions of ($p_s$, $r_{eh}$) and ($p_s$, $r_{gz}$), respectively.

In the FCFS scenario shown in Figure~\ref{fig:combined_aoi}(a), the average AoI increases sharply as the energy harvesting zone radius expands. This behavior aligns with previous observations, where the AoI rises with higher secondary user access probabilities. The increase can be attributed to the fact that a larger number of secondary users not only have sufficient energy to be active but are also allowed to transmit, thereby intensifying interference and causing longer service delays for the primary user.
In the QR scenario (Figure~\ref{fig:combined_aoi}(b)), the increase in average AoI is more moderate compared to the FCFS case, owing to the replacement mechanism that helps maintain fresher updates in the queue.
Finally, the GW policy (Figure~\ref{fig:combined_aoi}(c)) consistently outperforms the other two strategies, following a trend similar to that of the QR policy but yielding overall lower average AoI values.

Finally, Figure~\ref{fig:combined_aoi} illustrates the impact of the guard zone radius $r_{gz}$ and the secondary user access probability $p_s$ on the average AoI. In the FCFS scenario (Figure~\ref{fig:combined_aoi}(a)), the average AoI decreases rapidly as the guard zone expands. This behavior occurs because a larger guard zone prevents more secondary users from transmitting, thereby reducing interference at the primary receiver and consequently lowering the AoI.
For both the QR (Figure~\ref{fig:combined_aoi}(b)) and GW (Figure~\ref{fig:combined_aoi}(c)) policies, a similar decreasing trend in average AoI is observed with increasing $r_{gz}$, highlighting the consistent interference-mitigation benefits across all three queueing strategies.

\section{Conclusions}
The results highlight the critical influence of the energy harvesting zone radius ($r_{eh}$) and the guard zone radius around the primary receiver ($r_{gz}$) on overall system performance. While increasing $r_{eh}$ initially allows more STs to harvest energy and become active, excessive expansion ultimately degrades performance due to higher interference levels. Similarly, enlarging the guard zone $r_{gz}$ effectively mitigates interference at the PR and can significantly enhance AoI performance; however, this improvement comes at the expense of reduced transmission opportunities for secondary users, thereby decreasing their throughput.

Future work could investigate advanced EH models, adaptive access control, and extensions to heterogeneous network settings. 
Furthermore, integrating machine learning techniques to dynamically adjust zone radii and access policies represents a promising direction for optimizing the trade-off between throughput and information freshness in cognitive networks.

\section*{Acknowledgment}
This work has been supported in part by the Swedish Research Council (VR), ELLIIT, the European Union (6G-LEADER) under Grant 101192080; the European Union’s Horizon Europe Research and Innovation Programme under the Marie Skłodowska-Curie Grant under Agreement 101131481 (SOVEREIGN).

\bibliographystyle{IEEEtran}
\bibliography{ref}

\end{document}